\documentstyle[psfig]{mn}

\def\deg{^{\circ}}

\def\gtsim{\:{_>\atop{^\sim}}\:}

\def\teff{$T_{\rm{eff}}$}
\def\logg{$\log g$}

\def\1pos{$x_1,y_1$}
\def\2pos{$x_2,y_2$}
\def\3pos{$x_3,y_3$}
\def\4pos{$x_4,y_4$}

\def\cov{{\bf S}}

\def\xvec{{\bf x}}

\def\yvec{{\bf y}}

\def\uvec{{\bf u}}
\def\evec{\mbox{\boldmath $\varepsilon$}_p} 

\def\cmean{\overline{C^p}}

\def\sig68{$\sigma_{\rm{68}}$}
\def\sigrms{$\sigma_{\rm{rms}}$}
\def\sigint{$\sigma_{\rm{int}}$}

\title[Neural Network Classification of Stellar Spectra]
{Automated Classification of Stellar Spectra. II: 
Two-Dimensional Classification
with Neural Networks and Principal Components Analysis}
\author[C.A.L.\ Bailer-Jones et al.]
{Coryn A.L.\ Bailer-Jones$^1$\thanks{Present Address: 
Mullard Radio Astronomy Observatory,
Cavendish Laboratory, Madingley Road, Cambridge, 
CB3 0HE, UK}\thanks{email: calj@mrao.cam.ac.uk},
Mike Irwin$^2$, Ted von Hippel$^3$\\
$^1$ Institute of Astronomy, Madingley Road, Cambridge, CB3 0HA, UK\\
$^2$ Royal Greenwich Observatory, Madingley Road, Cambridge, CB3 0EZ, UK\\
$^3$  Department of Astronomy, University of Wisconsin, Madison, WI 53706, USA\\
}
\date{Submitted 7 September 1997}
\pubyear{1997}

\begin{document}

\maketitle

\begin{abstract}
We investigate the application of neural networks to the automation of
MK spectral classification.  The data set for this project consists of
a set of over 5000 optical (3800--5200\,\AA) spectra obtained from
objective prism plates from the Michigan Spectral Survey.  These
spectra, along with their two-dimensional MK classifications listed in
the Michigan Henry Draper Catalogue, were used to develop supervised
neural network classifiers. We show that neural networks can give
accurate spectral type classifications (\sig68 = 0.82 subtypes,
\sigrms = 1.09 subtypes) across the full range of spectral types
present in the data set (B2--M7). We show also that the networks yield
correct luminosity classes for over 95\% of both dwarfs and giants
with a high degree of confidence.

Stellar spectra generally contain a large amount of redundant
information. We investigate the application of Principal Components
Analysis (PCA) to the optimal compression of spectra. We show that PCA
can compress the spectra by a factor of over 30 while retaining
essentially all of the useful information in the data set.
Furthermore, it is shown that this compression optimally removes noise
and can be used to identify unusual spectra.

This paper is a continuation of the work done by von Hippel et~al.\
(1994) (Paper I)\nocite{vonhippel_94a}.
\end{abstract}

\begin{keywords}
methods: analytical, data analysis, numerical - stars: fundamental parameters
\end{keywords}


\section{Introduction}

The MK classification of stellar spectra (Morgan, Keenan \& Kellman
1943\nocite{morgan_43a}; Keenan \& McNeil 1976\nocite{keenan_76a};
Morgan, Abt \& Tapscott 1978\nocite{morgan_78a}) is an important tool
in stellar and galactic astrophysics.  In addition to providing
fundamental stellar information it was, for example, central to the
discovery of nearby Galactic spiral arms (Morgan, Sharpless \&
Osterbrock 1952; Morgan, Whitford \& Code
1953).\nocite{morgan_52a}\nocite{morgan_52a}

MK classification is usually performed by a trained expert visually
matching the overall appearance of a spectrum to the `closest' MK
standard spectrum.  Such a qualitative method of classification
suffers from subjective decisions and may differ from person to
person: what is deemed as `close' by one person may not be `close' for
another.  In addition, visual classification is very time consuming,
with an expert classifying a few $10^5$ stars in a dedicated lifetime.
Spectra collected from large spectral surveys, often as a by-product
of other surveys (e.g.\ the Sloan Digital Sky Survey (Kent
1994)\nocite{kent_94a}) will have to be classified by automated
means. Thus if stellar classification is to continue to be useful to
the astronomical community, it has to be made faster and put on a more
quantitative and objective basis.

In this paper we investigate the application of neural networks to the
MK classification of optical stellar spectra.  The so-called
`supervised' neural networks used in this project are implemented to
yield an accurate mapping between a data domain (the stellar spectra)
and a classification domain (the MK classifications).  While visual
classifiers have mentally determined this mapping, they have not
quantified it.  This mapping is, however, present intrinsically in a
large set of classified spectra.  The neural network's resultant
classification criteria will be essentially equivalent to the human's
criteria. However, whereas a human's criteria may vary from adverse
physiological and psychological factors such as health and mood, the
network will retain a consistent set of classification criteria.  We
will also demonstrate how the technique of Principal Components
Analysis (PCA) can be used to optimally compress the spectra.  This
has a number of advantages including the preferential removal of noise
and an ability to isolate bogus spectra. Furthermore, using
PCA-compressed spectra (rather than complete spectra) in the neural
network classifiers leads to reduced training times and better
convergence stability.

While MK classification will continue to be a useful tool to
astronomers, it becomes increasingly desirable to obtain physical
parameters (\teff, \logg, etc.)\ for stars. Bailer-Jones et~al.\
(1997b)\nocite{bailerjones_97b} describe a neural network approach to
the parametrization of stellar spectra by training a neural network
on synthetic spectra.

\section{Previous Classification Work}

There have been a number of attempts in the past to automate stellar
spectral classification.  Kurtz (1982)\nocite{kurtz_82a} classified
low (14\,\AA) resolution spectra using cross-correlation with standard
spectra and achieved a mean classification error of 2.2 spectral
subtypes for stars in the range B0 to M2.  The same technique gave
poor luminosity classification results.  LaSala
(1994)\nocite{lasala_94a} used the related technique of minimum
distance classification to classify a set of 350 B-star spectra, and
achieved a mean error of 1.14 spectral subtypes. 

The classification work of von Hippel et~al.\ (1994) (Paper
I)\nocite{vonhippel_94a} was one of the first applications of neural
networks to stellar spectral classification.  Their neural network
solution based on a set of 575 spectra gave an RMS classification
error of 1.7 spectral subtypes (and a 68-percentile error of 1.4
spectral subtypes) for spectra in the range B3 to M4.  Gulati et~al.\
(1994)\nocite{gulati_94a} trained a neural network on a set of 55
spectra giving an incomplete coverage of spectral classes O through to
M. While they reported classification errors of 2 subtypes, it should
be noted that they used a very complex neural network with over 18,000
free parameters (network weights), with no justification of why such a
complex network was required.  The result is that the determination of
these weights was likely to be poorly constrained by the small amount
of training data used.

There have also been attempts to classify spectra beyond the visual.
Weaver \& Torres-Dodgen (1995)\nocite{weaver_95a} used neural networks
to classify infrared spectra (5800\,\AA\ to 8900\,\AA) of A stars at
15\,\AA, and achieved spectral type and luminosity class
classification precisions of 0.4 subtypes and 0.15 luminosity classes
respectively. They have recently achieved good results in the infrared
for a wide-range of spectral types (O--M) and luminosity classes
(I--V) (Weaver \& Torres-Dodgen 1997)\nocite{weaver_97a}.  Vieira \&
Pons (1995)\nocite{vieira_95a} used a neural network trained on a set
of 64 IUE ultraviolet spectra (150\,\AA\ to 3200\,\AA) in the range O3
to G5, and reported a classification error of 1.1 spectral subtypes.
It was unclear, however, why a network with 110,000 weights was
required.

Whitney (1983)\nocite{whitney_83a} has examined the use of Principal
Components Analysis for spectral classification of a set of 53 A and F
stars. His data set consisted of 47 photoelectric measurements of
spectra over the wavelength range 3500\,\AA\ to 4000\,\AA.  He applied
PCA to his data set and then performed a regression on the three most
significant components, achieving an average classification error of
1.6 spectral subtypes.

\section{The Spectral Data}\label{data_red}

The classification techniques described in this paper were developed
using a set of 5000 spectra taken from the Michigan Spectral
Survey (Houk 1994)\nocite{houk_94a}. The data reduction method
is described in Paper I and in more detail in Bailer-Jones et~al.\ (1997a).
The present work expands the data set of Paper I by a factor of ten
and doubles the spectral resolution. The wavelength range is also slightly
different, with the details summarized in Table~\ref{plt_det}.
\begin{table}
\begin{center}
\caption{The spectral data.}
\begin{tabular}{|l|l|}\hline
Plate type      & IIaO \\
Plate size  & $\approx 20 \times 20$ cm \\
            & $\approx 5\deg \times 5\deg$ \\
            & 12,000 $\times$ 12,000 pixels \\
            & 289 Mb (FITS) \\
Plate scale & 96.62 arcsec ${\rm mm}^{-1}$ \\
Dispersion  & 108\,\AA/mm at H$\gamma$ \\
Scanning pixel size & 15$\mu$m \\
                    & $\Rightarrow$ 1.45 arcsec ${\rm pix}^{-1}$ \\
                    & $\Rightarrow$ 1.6\,\AA\ at H$\gamma$\\
                    & (1.05\,\AA\,${\rm pix}^{-1}$ @ 3802\,\AA \\
                    & 2.84\,\AA\,${\rm pix}^{-1}$ @ 5186\,\AA)\\
Coverage of final spectra & 3802--5186\,\AA \\
Magnitude limit of plates & B $\sim 12$ \\ \hline
\end{tabular}
\label{plt_det}
\end{center}
\end{table}
The classification information required to train and test the neural
networks is taken from the {\it Michigan Henry Draper} (MHD) catalogue
(Houk \& Cowley 1975; Houk 1978, 1982; Houk \& Smith-Moore 1988).
\nocite{houk_75a}\nocite{houk_78a}\nocite{houk_82a}\nocite{houk_88a}
In this paper we only examine the automated classification of normal
stars in terms of their MK spectral type and luminosity classes.
However, the MHD contains considerable additional information,
particularly with regard to peculiarities, so this catalogue would be
suitable for developing more detailed automated classifiers.

\begin{figure}
\centerline{
\psfig{figure=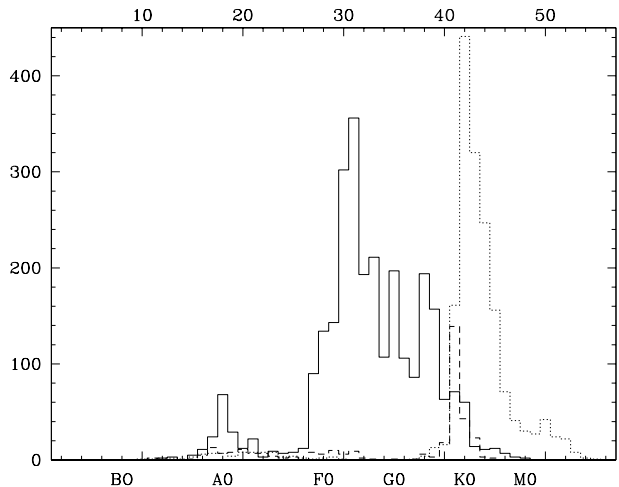,width=0.5\textwidth,angle=270}
}
\caption{Distribution of spectral types for each luminosity class 
in data set B. The dotted line represent
giants (III), the dashed line subgiants (IV) and the solid line dwarfs (V).
The distribution for data set A is very similar.
}
\label{dist_B}
\end{figure}

Our set of 5144 spectra contains stars over a wide range of spectral
types (B2--M7) for luminosity classes III, IV and V as well as the
`intermediate' luminosity classes III/IV and IV/V. This set, hereafter
referred to as data set `A', was used to develop a spectral type
classifier.  A second data set, `B', contains only `whole' luminosity
classes (i.e.\ not the III/IV and IV/V spectra). This set of 4795
spectra is used to develop the luminosity class classifier.  The
distribution of spectral types in this latter set is shown in
Figure~\ref{dist_B}. The spectra were normalized to have
equal areas, i.e.\ equal total intensities, thus removing any scale
differences resulting from different apparent magnitudes.  Line-only
spectra were obtained for both data sets, using a non-linear
rectification method (Bailer-Jones~et~al.\ 1997a). In this paper we
investigate automated classification with both line-only
and line+continuum spectra.

\section{Neural Network Models}\label{ann}

A neural network is a computational tool which will
provide a general, non-linear parameterized mapping between a set of
inputs (such as a stellar spectrum) and one or more outputs (such as a
spectral classification).  The type of neural network architecture
used in this paper is known as a {\it multi-layer perceptron} neural
network (e.g.\ Hertz, Krogh \& Palmer 1991\nocite{hertz_91a}; Bishop
1995\nocite{bishop_95a}; Lahav et~al.\ 1996).  In order to give the
correct input--output mapping, the network is trained on a set of
representative input--output data. Training proceeds
by optimising the network parameters (the `weights') to give the
minimum classification error.  With the weights fixed, the network is
used to produce outputs (MK classifications) for unclassified inputs
(stellar spectra), effectively by interpolating the training data.
Note that the output from a neural network is some non-linear function
of {\it all} of the network inputs.  Thus the network's
classifications are based on the appearance of the {\it whole}
spectrum: we do not have to tell the network in advance which spectral
lines are relevant.

Network training used the methods of gradient descent and
backpropagation (Rumelhart, Hinton \& Williams 1986).
Network performance was found to be insensitive to
the exact vaules of the `gain' and `momentum' parameters.  It was
determined that 1000 training iterations were sufficient to ensure
that the network error, as evaluated on an independent test data set,
had reached its minimum error. Up to 50 times as many iterations gave
only negligible improvement.  Training a neural network on 2500
spectra represented as 50 PCA coefficients typically took about an
hour. The application of these trained networks to then classify a
similar number of spectra is a few seconds.  Further details can be
found in Bailer-Jones (1996).

Spectral type classification was performed by representing the 57 MK
classifications in the MHD as points on a continuous scale of numbers
1--57 (Table~\ref{codes1_tab}).
This is reasonable
as we know that spectral type is closely related to effective
temperature (\teff) and MK spectral types are essentially binnings of
a continuous sequence. The
appropriate neural network therefore had a single output giving a
continuous number in the range 1--57. We shall refer to this as {\it
continuous mode}.  Note that although the network is trained on
integer or half-integer values, it can produce any real-value
classification for new spectra.
\begin{table}
\begin{center}
\caption{Numerical coding of the MK spectral types.
`SpT' will be used to label this code, so that 44 SpT $\equiv$ K2.
The MHD catalogue omits some classes (e.g.\ F4 and G7).}
\begin{tabular}{|ll||rl||rl|}\hline
\hspace{1em} 1   &   O3\hspace{2em}  &\hspace{1em} 18  & A0\hspace{1em} & \hspace{1em}38  & G3  \\
\hspace{1em} 2   &      O4  &   19  &   A1  &   39  &   G5  \\
\hspace{1em} 3   &      O5  &   20  &   A2  &   40  &   G6  \\
\hspace{1em} 4   &      O6  &   21  &   A3  &   41  &   G8  \\
\hspace{1em} 5   &      O7  &   22  &   A4  &   42  &   K0  \\
\hspace{1em} 6   &      O8  &   23  &   A5  &   43  &   K1  \\
\hspace{1em} 7   &      O9  &   24  &   A6  &   44  &   K2  \\
\hspace{1em} 7.5 &      O9.5 &  25  &   A7  &   45  &   K3  \\
\hspace{1em} 8   &      B0   &  26  &   A8  &   46  &   K4  \\
\hspace{1em} 8.5 &      B0.5 &  27  &   A9  &   47  &   K5  \\
\hspace{1em} 9   &      B1  &   28  &   F0  &   48  &   M0  \\
\hspace{1em} 10  &      B2  &   29  &   F2  &   49  &   M1  \\
\hspace{1em} 11  &      B3  &   30  &   F3  &   50  &   M2  \\
\hspace{1em} 12  &      B4  &   31  &   F5  &   51  &   M3  \\
\hspace{1em} 13  &      B5  &   32  &   F6  &   52  &   M4  \\
\hspace{1em} 14  &      B6  &   33  &   F7  &   53  &   M5  \\
\hspace{1em} 15  &      B7  &   34  &   F8  &   54  &   M6  \\
\hspace{1em} 16  &      B8  &   35  &   G0  &   55  &   M7  \\
\hspace{1em} 17  &      B9  &   36  &   G1  &   56  &   M8  \\
\hspace{1em} 17.5 &     B9.5 &  37  &   G2  &   57  &   M9  \\ \hline
\end{tabular}
\label{codes1_tab}
\end{center}
\end{table}

For the luminosity class problem, we used a network in {\it
probabilistic mode}. This refers to a network with several outputs,
each output referring to a mutually exclusive class. In our case we
had three outputs, with one output corresponding to each of classes
III, IV and V. In this mode, the values in each node can be
interpreted as the probability that the input spectrum is of that
particular luminosity classes.  Probabilistic interpretation of neural
networks in both probabilistic mode (also referred to in the neural
network literature as `classification') and continuous mode (also
referred to as `regression') can be taken much further. In particular,
the outputs can be interpreted as Bayesian posterior probabilities
(e.g.\ Richard \& Lippmann 1991; MacKay 1995).

\subsection{Committee of Networks}\label{multi_nets}

Identical neural networks trained from different initial random
weights should ideally converge on the same weights and hence produce
identical input--output mappings.  However, given the high
dimensionality and complexity of the error surface which is
explored during training, it is unlikely that numerical
minimization procedures with different initializations would converge
on exactly the same final weight vector.

We can reduce the effects of this `convergence noise' problem by using
a {\it committee of neural networks}.  The committee consists of
$L$ identical networks which are separately trained from
different initial weights.  When the network is used in continuous
mode, the committee classification of the $p^{th}$ spectrum, $\cmean$,
is just the average of the individual network classifications.  In
probabilisitic mode, the outputs from each network corresponding to a
given class are summed to give the (unnormalized) committee
probability of that class.  In some applications, $\cmean$ will be a
more accurate classification than any of the individual network
classifications (Bishop 1995). All classification results presented in
this paper were obtained with a committee of ten neural networks.

\subsection{Neural Network Error Measures}\label{err_meas}

The performance of a trained neural network (or committee thereof) is
evaluated by comparing its classifications of an independent set of
spectra with their `true' classifications in the MHD catalogue. We
must not evaluate the performance of the neural network using the data
on which it was trained. This is because it is possible for the
network to overfit the training data rather than capture the
underlying input--output mapping which the training data
represent. This can occur if there is either insufficient data to
constrain the determination of the network weights, or if the training
data is not representative of the problem in hand. Such an overfitted
network would typically produce very low classification errors on the
training data yet produce relatively large errors on an independent
test data set.  Our procedure was therefore to train a network on
half of the spectra and test its performance on the other half,
there being approximately 2500 spectra in each half.

We use the following error measures to evaluate the performance of our
networks. The first is the RMS error, \sigrms, of the difference
between the network classifications and the `true' classifications.
This statistic suffers from the usual problem that it is sensitive to
outliers and may not, therefore, be very characteristic of the
majority of residuals in the core of the distribution.  A more robust
measure uses only the central $68\%$ of the residuals, \sig68. If the
residuals are distributed as a Gaussian, \sig68\ is the 1$\sigma$
standard deviation of a Gaussian distribution.  Both \sigrms\ and
\sig68\ are {\it external} errors, because they are measured with
respect to a set of ideal classifications which are external to the
neural network. 

Due to the `convergence noise' problem (section~\ref{multi_nets}), a
network re-trained from different initial weights would give slightly
different classifications. This level of difference is characterised
by the {\it internal} error and is evaluated using the
committee of networks. The internal error for a single spectrum is:
\begin{equation}
\sigma_{int}^p = \sqrt{ \frac{1}{L-1}\sum_{l=1}^{l=L}(C_l^p - \cmean)^2 } \ \ .
\end{equation}
$C_l^p$ is the classification given by the $l^{th}$ network in the
committee. The total internal error, \sigint, is $\sigma_{int}^p$
averaged over all spectra, and can be considered as the contribution
to the total (external) error on account of imperfect network
convergence.

An additional consequence of the `convergence noise' is that a single
value of the external error, \sig68, is not exact. Thus if we wish to
compare values of \sig68\ produced by different network models, then
we need to know the uncertainty in \sig68\ in order to known whether
the difference between two values of \sig68\ is statistically
significant.  A suitable measure of this uncertainty is given by the
standard error in the $\sigma$ of a Gaussian distribution,
\begin{equation}
\varepsilon = \frac{\sigma_{\rm{68}}}{\sqrt{2M}} \ \ ,
\label{err2}
\end{equation}
where $M$ is the number of spectra in the test set. This result
holds exactly in the limit as $M \rightarrow \infty$.

\begin{figure}
\centerline{
\psfig{figure=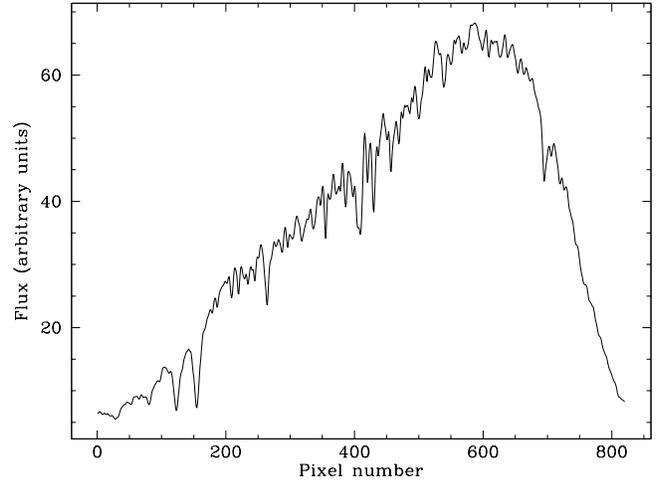,angle=270,width=0.5\textwidth}
}
\caption{Average spectrum for line+continuum version 
of spectra in data set A.
}
\label{avspec2}
\end{figure}

\section{Principal Components Analysis}

It is usually desirable -- and often essential -- to reduce the
dimensionality of a data set prior to classification.  Dimensionality
reduction often leads to enhanced reliability when using neural
networks to give a generalized mapping, on account of the reduced
number of parameters in the network. This will also lead to greatly
reduced training times. While classification with complete spectra can
produce good results (Paper I), dimensionality reduction may be
essential in some applications, such as when data transmission rates
 from space-based observatories are limited (e.g.\ Lindegren \&
Perryman 1996).

Principal Components Analysis (PCA) is one method for achieving a
dimensionality reduction.  PCA is a method of representing a set of
$N$-dimensional data by means of their projections onto a set of $r$
optimally defined axes.  As these axes (the {\it principal
components}) form an orthogonal set, PCA yields a {\it linear}
transformation of the data.  A compression of the data is obtained by
ignoring those components which represent the least variance in the
data.  The compressed spectra, as represented by their projections,
onto the most significant principal components, are then used as the
neural network inputs.  In this section we will demonstrate the
benefits of a PCA preprocessing of spectra, such as noise removal and
identification of bogus spectra, and highlight some of its problems.
In the next section we shall demonstrate that high quality
classifications can be achieved with these compressed spectra.

Below we discuss the application of PCA to the line+continuum spectra
of data set A. A separate analysis was carried out for the line-only
spectra (Bailer-Jones 1996).  PCA has been used in several areas
of astronomy, including stellar spectral classification (Deeming
1964\nocite{deeming_64a}; Whitney 1983\nocite{whitney_83a};
Storrie-Lombardi et~al.\ 1994\nocite{storrie-lombardi_94a}), galaxy
spectral classification (Folkes, Lahav \& Maddox
1996\nocite{folkes_96a}) and quasar spectral classification (Francis
et~al.\ 1992)\nocite{francis_92a}. Further details of the technique
can be found in texts (e.g.\ Murtagh \& Heck 1987).

\subsection{The Principal Components}\label{eig_anal}

\begin{figure*}
\begin{minipage}{14cm}
\centerline{
\psfig{figure=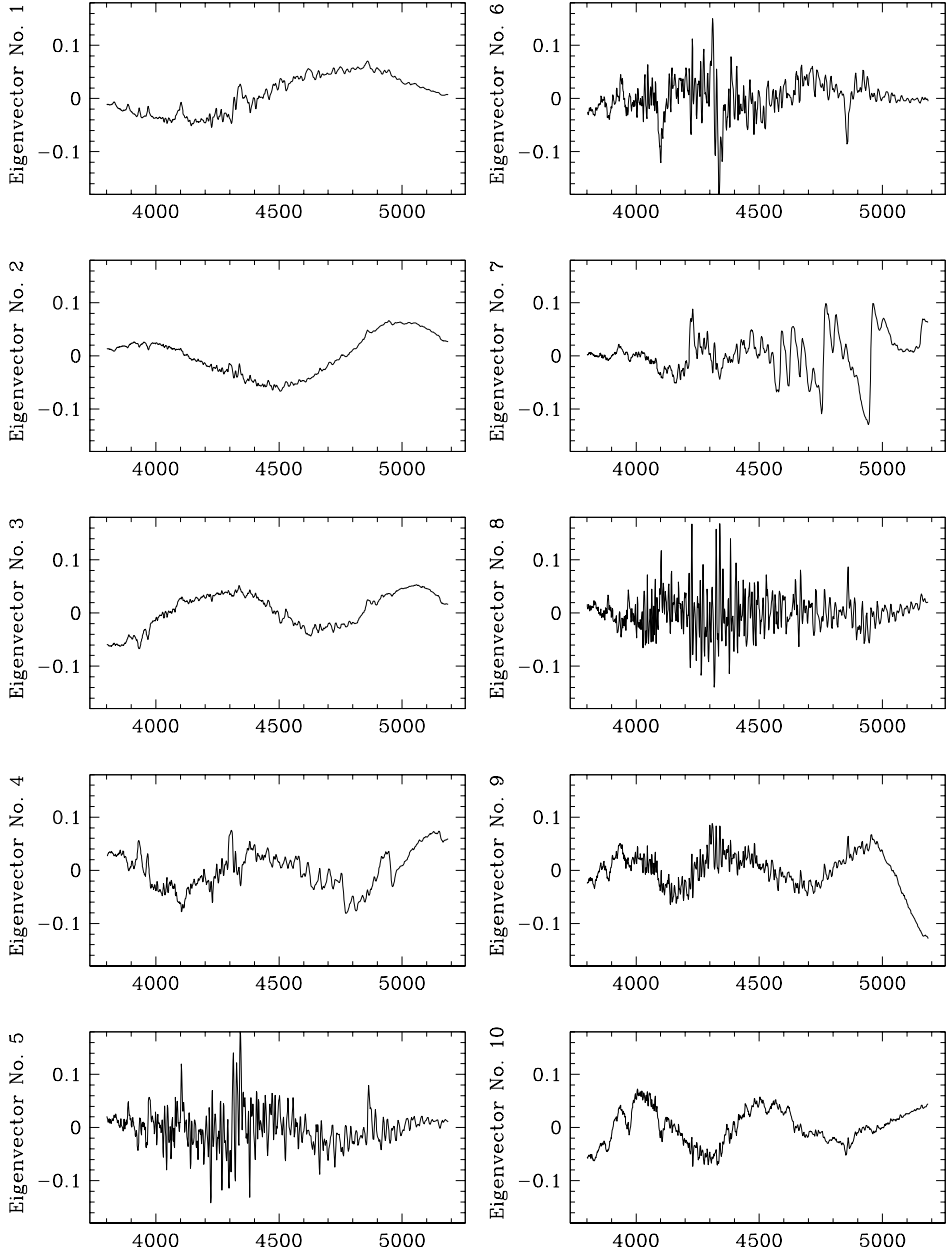,angle=0,width=0.90\textwidth}
}
\caption{The first ten principal components 
of the line+continuum spectra in data set A. The principal components
are normalized eigenvectors plotted against wavelength.
A number of spectral features can be seen.
In particular, the $7^{th}$ eigenvector strongly represents
the TiO bands of late-type stars.}
\label{eigenvec1}
\end{minipage}
\end{figure*}

PCA was performed on zero-mean spectra: the subtracted average spectrum, 
$\overline{\xvec}$, is shown in Figure~\ref{avspec2}.
Figure~\ref{eigenvec1} shows the first ten
principal components, $\uvec$, plotted on a common vertical scale.  Because the
principal components are eigenvectors of a symmetric matrix, they are
orthogonal, and it is convenient also to normalize them to unit
length, so that $\uvec_i^T.\uvec_j = \delta_{i,j}$. As the
eigenvectors are simply rotations in the $N$-dimensional data space of
the original axes on which the spectra are defined, they resemble
spectra, in particular in that they have the same number of elements
(820) as the original spectra.

It is interesting that both the stellar continuum and the individual
spectral lines are distributed across many eigenvectors. For example,
the Ca II H\&K lines at 3934\,\AA\ and 3969\,\AA\ are distinct in the
first four eigenvectors as well as the average spectrum.  (Note that
the sign of the eigenvectors is arbitrary, as the admixture
coefficients -- the projections of the spectra onto the principal
components -- can be negative.)  That the features do not separate
into different components is not surprising: We known from the physics
of line formation in stellar photospheres that a spectrum is not a
linear combination of spectral features, so we should not expect a
linear decomposition of the spectrum (such as PCA) to clearly isolate
these spectral features.  Features are generally distributed across
many components, e.g.\ components 5 and 8 which show many lines common
to a wide range of spectral types. However, some features are
predominantly represented by a single components.  For example, it can
be seen that the TiO bands (which extend redward from about 4500\,\AA)
characteristic of M stars are more strongly represented in the
$7^{th}$ principal component than any other component.

The principal components represent sources of variance in the
data. Thus the most significant principal components show those
features which vary the most between the spectra: it is important to
realise that the principal components do not simply represent strong
features. Note also that the eigenvectors obtained from PCA are
entirely dependent on the data. Therefore the eigenvectors for a
different set of stellar spectra are likely to be rather different.

\subsection{The Admixture Coefficients}\label{admix_sec}

\begin{figure*}
\begin{minipage}{14cm}
\centerline{
}
\caption{(This figure is provided as a separate GIF file.)
The first ten principal component admixture coefficients
for all spectra (line+continuum format) 
in data set A plotted against spectral subtype (SpT).
The correlation between the $7^{th}$ coefficient 
and spectral subtype for SpT $\gtsim 48$ 
(M stars) is
accountable by reference to Figure~\ref{eigenvec1} where we see that
the $7^{th}$ eigenvector gives a strong representation of the TiO
features in M stars. In general, however, the admixture
coefficients cannot be used individually to give spectral type
classifications.}
\label{spt_coefs1}
\end{minipage}
\end{figure*}
The projection of the $p^{th}$ spectrum onto the $k^{th}$ principal
components is known as the {\it admixture coefficient}, $a_{k,p}$.
Because PCA is only a linear transformation of the spectra, one would
not expect there to be a strong correlation between the stellar
classification parameters and the admixture coefficients.
Figure~\ref{spt_coefs1} shows the admixture coefficients for the
line+continuum spectra plotted against the coded MK spectral type,
SpT, for each of the first ten eigenvectors shown in
Figure~\ref{eigenvec1}.  No single coefficient shows a strong
correlation across the full range of subtypes, so classification
cannot be achieved using any one coefficient.  Some coefficients do,
however, show correlations over part of the spectral range. We saw in
Figure~\ref{eigenvec1} that the $7^{th}$ eigenvector represents some
features of late-type stars and we see in Figure~\ref{spt_coefs1}
that the corresponding admixture coefficient, 
shows a trend with spectral type for SpT~$\gtsim 48$.

\subsection{PCA as Data Compression and Noise Filter}\label{PCA_comp}

\begin{figure}
\centerline{
\psfig{figure=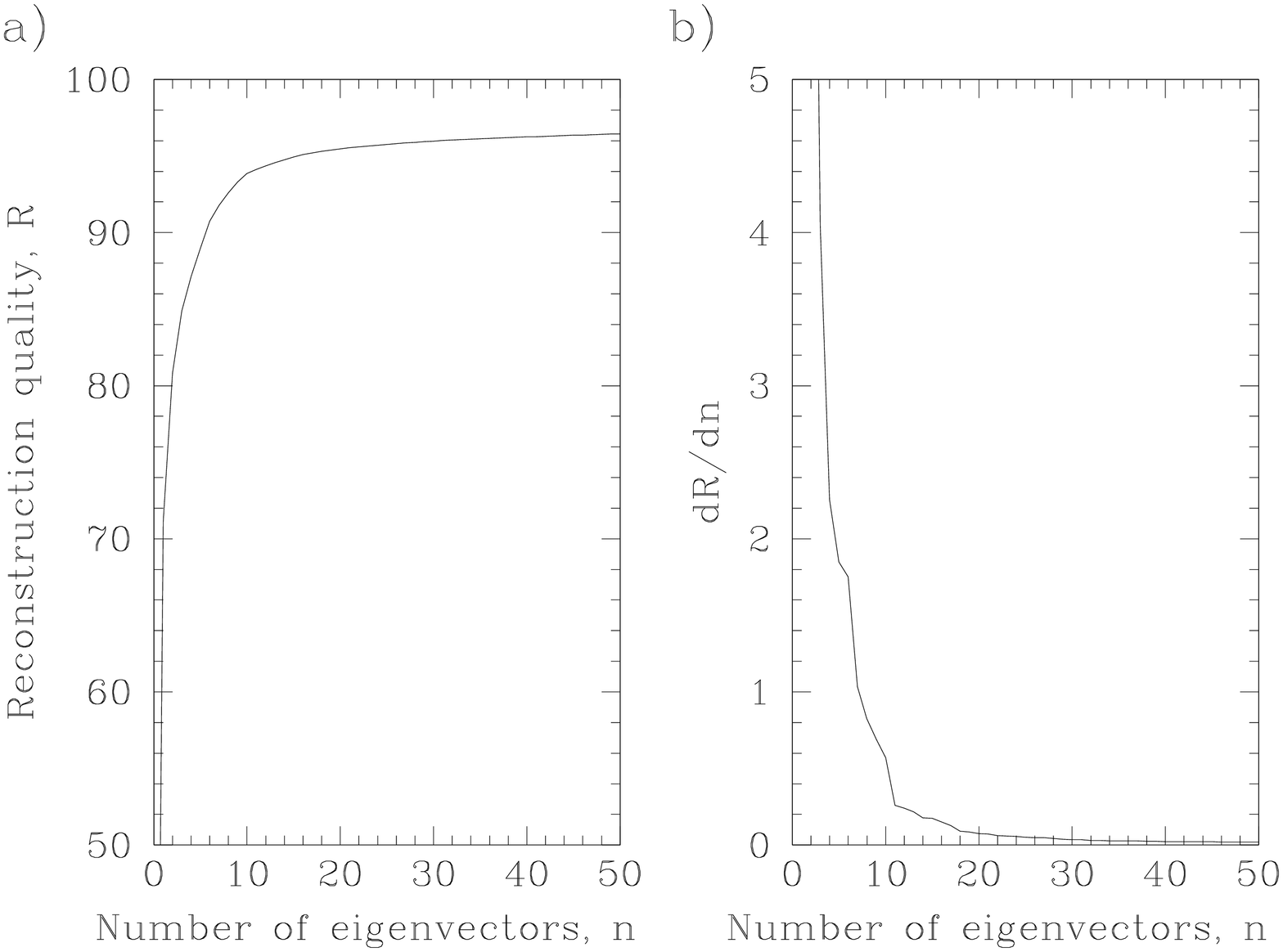,width=0.5\textwidth,angle=270}
}
\caption{Quality of spectral reconstruction with a reduced number of
eigenvectors. a) The quality of reconstruction increases dramatically
over the first 25 or so eigenvectors. b) At $n \approx 25$,
$\frac{dR}{dn} \approx 0$, and the remaining eigenvectors are
predominantly noise. Thus an optimal reconstruction only requires 
approximately the
first 25 eigenvectors.  $R$ is the
reconstruction error predicted by the eigenvalues
(equation~\ref{rec_err}) and so is the reconstruction error for the
whole data set.}
\label{reconerr_whole}
\end{figure}

The most significant principal components contain those features which
are most strongly correlated in many of the spectra. It follows that
noise -- which is uncorrelated with any other features by
definition -- will be represented in the less significant components.
Thus by retaining only the more significant components to represent
the spectra we achieve a data compression that preferentially removes
noise. The {\it reduced reconstruction}, $\yvec_p$, of the $p^{th}$ spectrum
$\xvec_p$, is obtained by using only the first 
$r$ principal components to reconstruct the spectrum, i.e.\
\begin{equation}
\yvec_p = \overline{\xvec} + \sum_{k=1}^{k=r}a_{k,p}\uvec_k \ \ ,
\ \ \ \ \ \ \ \ \ r < N \ .
\label{singfish}
\end{equation}
Let $\evec$ be the error incurred in using this reduced reconstruction.
By definition, $ \xvec_p = \yvec_p + \evec$, so
\begin{equation}
\evec = \sum_{k=r+1}^{k=N} a_{k,p}\uvec_k \ \ .
\end{equation}
Averaging over all spectra gives rise to the average error, ${\cal E}$,
 from which we define the figure-of-merit of the reconstruction 
quality of the whole data set as $R = 1 - {\cal E}$, and
\begin{equation}
R = 100\%\frac{\sum_{k=1}^{k=r} \lambda_k}{\sum_{k=1}^{k=N} \lambda_k} \ \ ,
\label{rec_err}
\end{equation}
where $\lambda_k$ is the $k^{th}$ eigenvalue of the covariance matrix,
$\cov$, of the data (Bailer-Jones 1996).  
Figure~\ref{reconerr_whole}a shows how $R$
varies with the number of eigenvectors used to reconstruct the
spectra, and Table~\ref{reconerr_tab} tabulates some of these values.
We see that only 25 eigenvectors ($\sim 3\%$ of the total) are
sufficient to reconstruct 95.8\% of the variance in the data. This
large factor of data compression is a great benefit for any approach
to the classification problem as it corresponds to a large reduction
of the dimensionality of the space required to describe the data. 
\begin{table}
\begin{center}
\caption{Selected values from Figure~\ref{reconerr_whole}.}
\begin{tabular}{|c|r|r|r|}\hline
  Number of  &             &         & \\ 
Eigenvectors & $R (= 1-E)$ & $dR/dn$ & $d^2R/dn^2$ \\ \hline
  1 &   71.135  &      --  & -- \\
  2 &   80.840  &   9.705  & $-$61.430 \\
  3 &   84.884  &   4.044  &  $-$5.661 \\
  4 &   87.139  &   2.255  &  $-$1.789 \\
  5 &   88.989  &   1.850  &  $-$0.405 \\
 10 &   93.861  &   0.570  &  $-$0.119 \\
 15 &   94.930  &   0.174  &  $-$0.004 \\
 20 &   95.458  &   0.074  &  $-$0.012 \\
 25 &   95.760  &   0.052  &  $-$0.005 \\
 30 &   95.972  &   0.036  &  $-$0.002 \\
 35 &   96.121  &   0.027  &  $-$0.001 \\
 40 &   96.244  &   0.023  &  $-$0.001 \\
 45 &   96.352  &   0.021  &  $-$0.000 \\
 50 &   96.449  &   0.018  &  $-$0.001 \\
820 &  100.000  &   0.000  &   0.000 \\ \hline
\end{tabular}
\label{reconerr_tab}
\end{center}
\end{table}

\begin{figure}
\centerline{
\psfig{figure=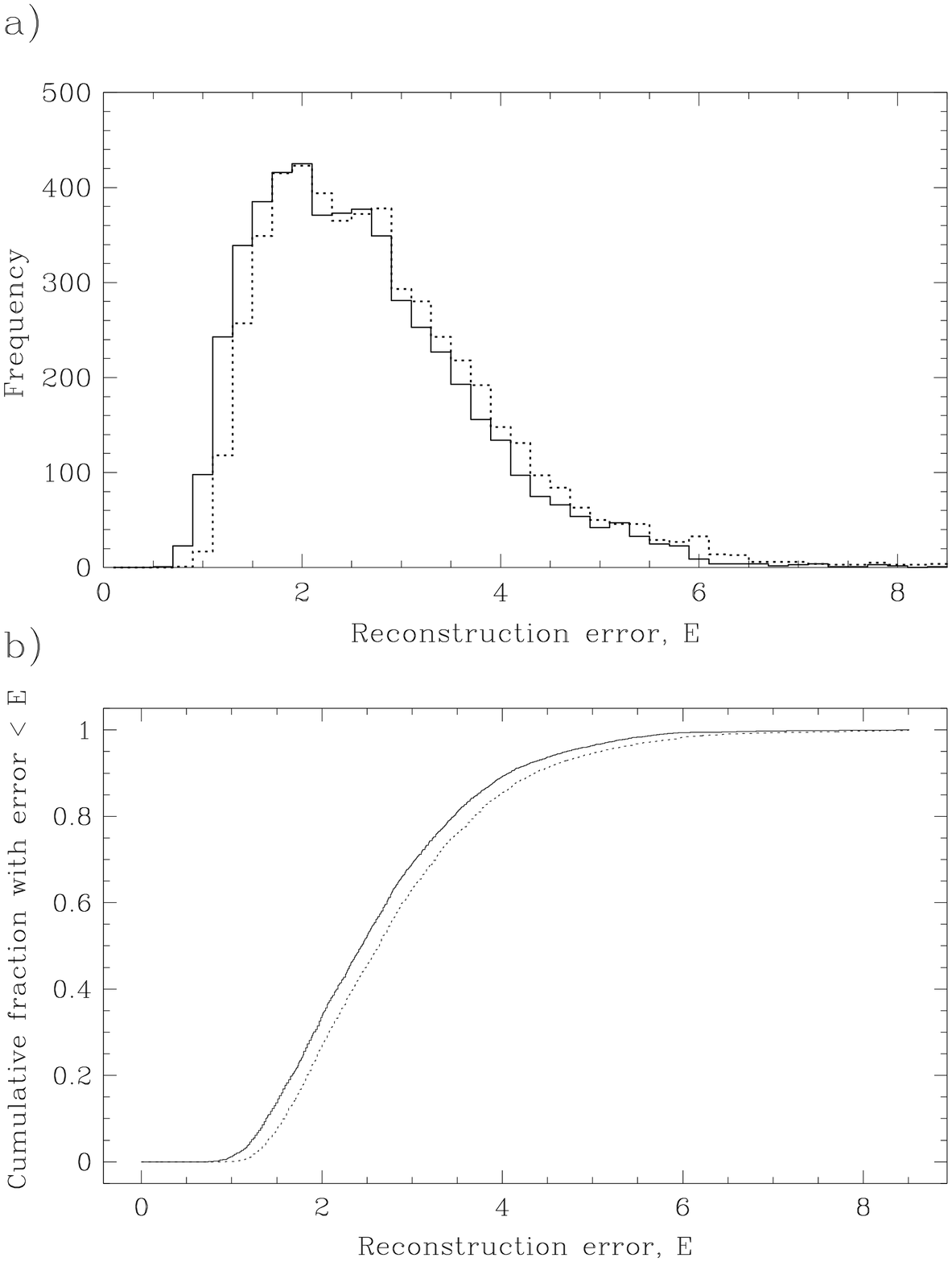,angle=0,width=0.5\textwidth}
}
\caption{Frequency distribution of the empirical reconstruction error
defined by equation~\ref{emp_recerr}.  The solid line shows the errors
for a 50-component reconstruction and the dashed line for a
25-component reconstruction.  (a) histogram of the reconstruction
errors.  (b) the cumulative distribution of (a).  This shows the
fraction of spectra which are reconstructed with an error less than
that shown on the horizontal axis. For example, 95\% of the spectra have $E
< 4.6\%$.}
\label{reconerr}
\end{figure}

It is convenient to define an empirical measure of the reconstruction
error for individual spectra
\begin{equation}
E = \frac{100\%}{S} \sum_{i=1}^{i=N} | y_{i,p} - x_{i,p} | \ \ ,
\label{emp_recerr}
\end{equation}
where $S$ is the total area under each spectrum, which was fixed
to a constant value when the spectra were area normalized.
This measure is useful as it does not require the existence of
eigenvalues for its evaluation, and hence can be used to compare
reconstruction errors between any data compression techniques.
The frequency distribution of these errors is shown in 
Figure~\ref{reconerr}.

\begin{figure}
\centerline{
\psfig{figure=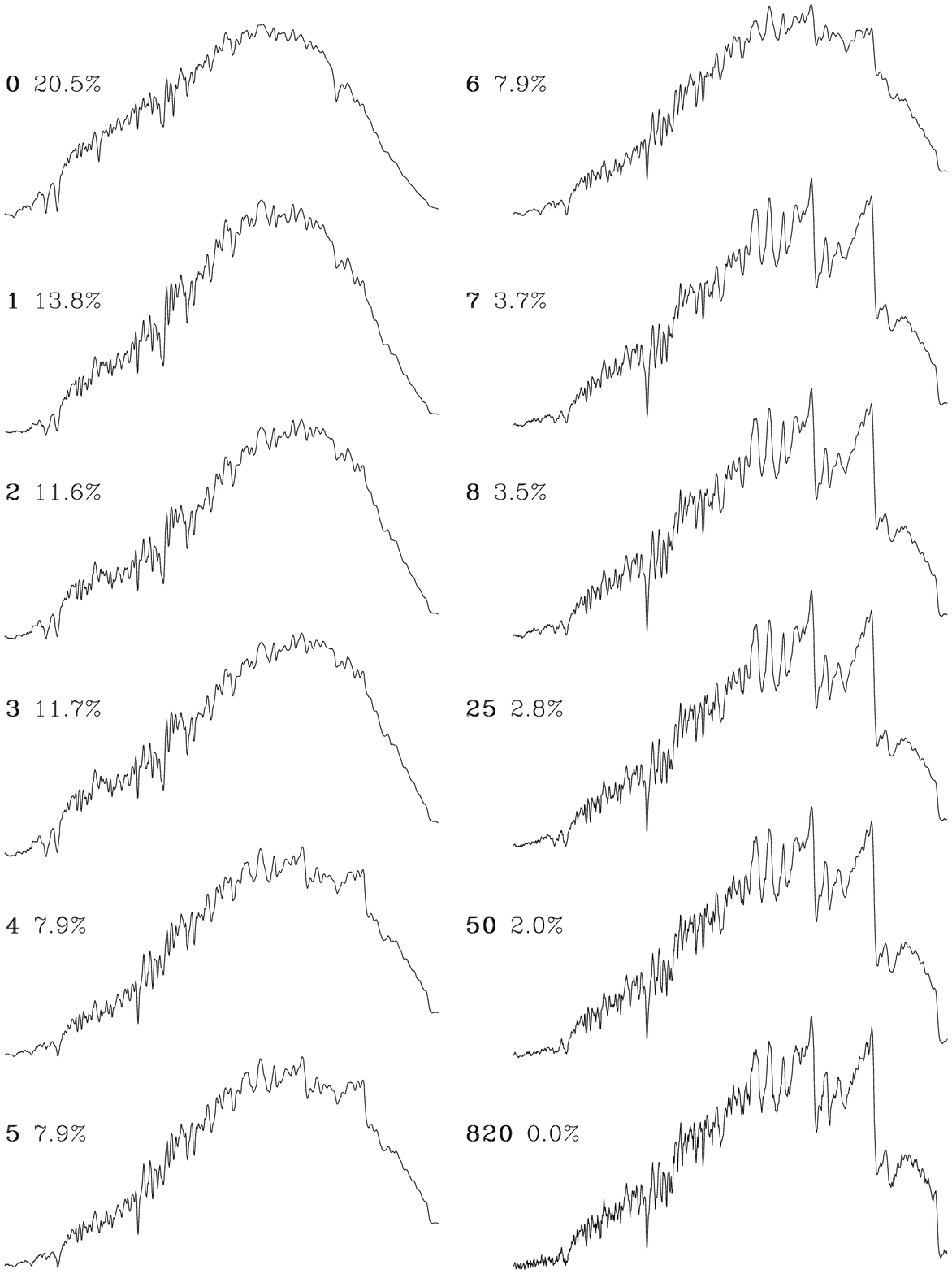,angle=0,width=0.5\textwidth}
}
\caption{Reconstruction of an M star (HD 14002, type M3/4 III, magnitude 9.4).
The figures in bold refer to the number of eigenvectors used to
reconstruct the spectrum. The percentage figures are the corresponding
reconstruction error, $E$. Note the improved reconstruction as the
$7^{th}$ component is added.}
\label{Mstar_recon}
\end{figure}
Figure~\ref{Mstar_recon} gives a visual presentation of spectral
reconstruction by showing an M~star spectrum reconstructed with an
increasing number of components. We have seen in
Figure~\ref{eigenvec1} that the $7^{th}$ eigenvector is representative
of the TiO bands in late type stars. In reconstructing this M~star
spectrum we see that the reconstruction error drops significantly when
the $7^{th}$ component is added, and that visually this $r=7$
spectrum is greatly improved over the $r=6$ one.

\begin{figure}
\centerline{
\psfig{figure=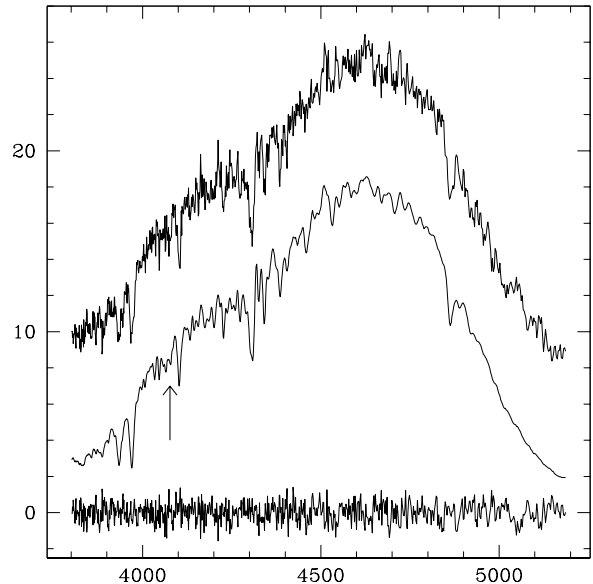,width=0.45\textwidth,angle=0}
}
\caption{Reconstruction of a faint G~star (HD 219795, G3 V, magnitude 11.1).
The reconstruction error is $E = 4.26\%$.
The top spectrum is the original
spectrum, the middle is the spectrum reconstructed with 25 components
and the bottom is the residual spectrum, i.e.\ reconstructed minus
original. The arrow marks the location of a weak luminosity sensitive
line Sr~II which is retained in the reconstruction. 
This spectrum is one of the faintest
(and hence noisiest) in the data set, so most reconstructions
are considerably better.}
\label{fGstar_diff}
\end{figure}

An optimal trade-off between compression (and noise removal) and
accurate spectral representation is achieved for $r=n$ when
$\frac{dR}{dn} \approx const$.  If a PCA were performed on a data set
of pure noise, no component would be a greater discriminant than
another, and their ranking would be random giving $\frac{dR}{dn} =
const$ for all $n$.  Turning this argument around, the point where the
$R$-$n$ plot levels off ($\frac{dR}{dn} \approx const$) is where the
components begin to be dominated by noise. This occurs between $n=20$
and $n=30$ (Figure~\ref{reconerr_whole}b). Figure~\ref{fGstar_diff}
compares a faint spectrum reconstructed with 25 components with the
original spectrum.  The reconstructed spectrum is considerably less
noisy and the residual spectrum contains no major features.
Note that if we had lower S/N data,
$\frac{dR}{dn}$ would turn constant at a smaller value of $N$.  Thus for
lower S/N data, the optimality criterion translates into retaining
fewer components.

\begin{figure}
\centerline{
\psfig{figure=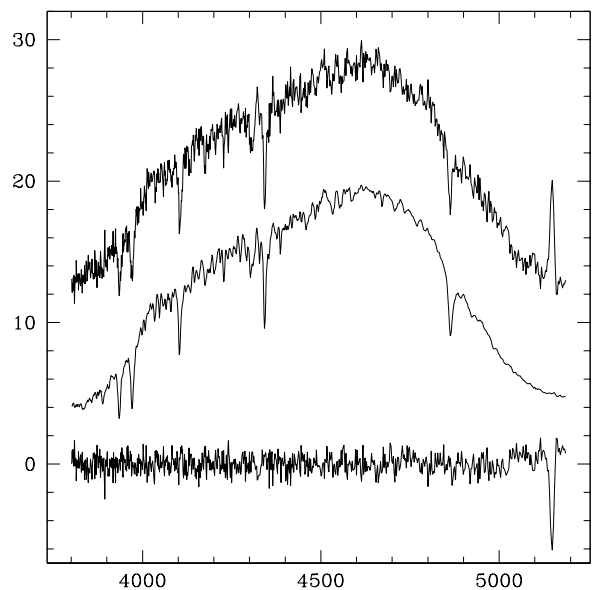,width=0.45\textwidth,angle=0}
}
\caption{The use of PCA in filtering out bogus features.
(See caption to
Figure~\ref{fGstar_diff}.)
The contaminating feature here is
probably due to a piece of dust on the plate during plate
scanning. The star is HD~3391, type F5~V, magnitude 10.8. The reconstruction
error is $E = 5.30\%$.}
\label{Dstar_diff}
\end{figure}

One of the drawbacks of PCA is that very weak spectral features or
features which are only present in a small fraction of the data will
be lost in a reduced reconstruction. This is because they show very
little correlation across the data set.  Thus the residual spectrum
will contain, in addition to noise, some weak features which are not
well correlated across the spectra.  However, as can be seen from
Figure~\ref{fGstar_diff}, by no means are all such features are lost.
Thus we see that the principal components (like neural network
classifiers) are sensitive to the relative frequency of occurrence of
features in the data set. An advantage of this is that PCA can be used
to filter out bogus features (e.g.\ plate scratches, strong cosmic
rays) because such features are rare and randomly positioned. This
is demonstrated in Figure~\ref{Dstar_diff}.


\subsection{Constructing New Spectra}

Having performed PCA on a set of stellar spectra (the {\it
construction set}), we may want to find the admixture
coefficients for a new set of data, such as more recently acquired
spectra. We do not have to re-evaluate the principal components using
the combined data sets: instead we can project the new spectra onto the
old components.  We can even obtain the admixture coefficients for
incomplete spectra, thus permitting a complete reconstruction. This
would be useful if we wanted to apply the neural network classifier to
new spectra with a slightly different wavelength coverage.

\begin{figure}
\centerline{
\psfig{figure=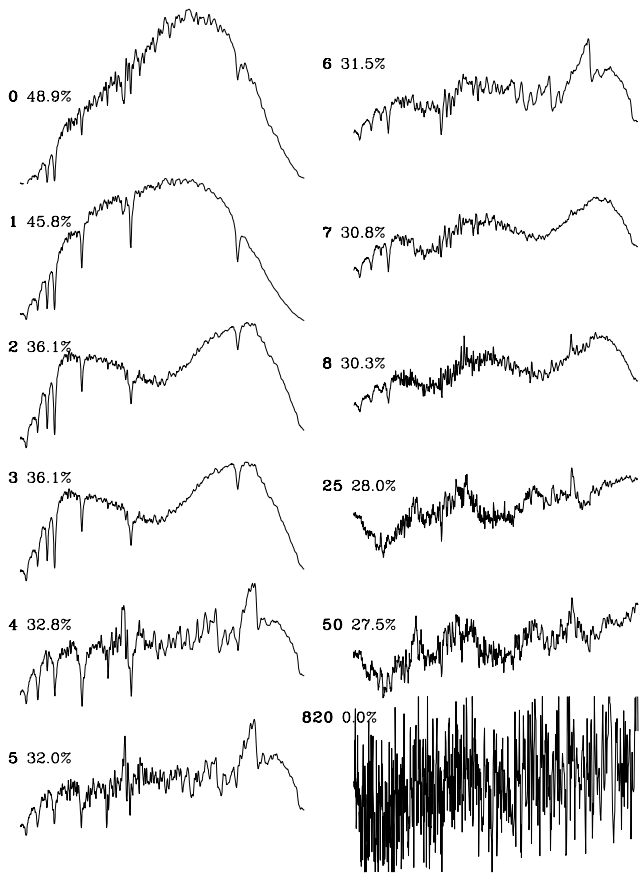,angle=0,width=0.5\textwidth}
}
\caption{Bogus spectra (in this case a patch of sky) are reconstructed with a
large error, which can be used to filter them out of the data set
prior to classification.  The figures in bold refer to the number of
eigenvectors used to reconstruct the spectrum. The percentage figures
are the corresponding reconstruction error, E.}
\label{Noise_recon}
\end{figure}

Another advantage of PCA is that the reconstruction error, $E$, can be
used to filter-out bogus and non-stellar spectra, as in such cases the
reconstruction error would be larger than the typical values in
Figure~\ref{reconerr}. As an example, Figure~\ref{Noise_recon} shows
how this filtering works by attempting to reconstruct sky noise.  The
25-component reconstruction gives an error of 28\%, which is several
times larger than the maximum reconstruction error of the spectra
in the construction set.

This method of rejection assumes that the data used to define the
principal components are representative of the stellar spectra which
we want to classify. Thus rare types of stars with strong features,
e.g.\ Me stars, would be filtered out along with all the bogus
spectra. Note that a neural network applied to either the complete
spectra or the PCA compressed spectra suffers from a similar problem:
spectral types which are relatively rare in the training data set will
be poorly classified.  But with PCA there is no reason why such
objects have to be blindly rejected: A large reconstruction error
indicates an unknown object, so PCA could be used as a coarse
front-end classifier to a more refined classification system.

\section{Spectral Type Results}\label{spt_class}

Neural networks were applied to the spectral type problem in
continuous mode (see section~\ref{ann}).  Each spectrum is represented
by the admixture coefficients of the first 25 or 50 principal
components. Data set A is used throughout.

\begin{figure}
\centerline{
\psfig{figure=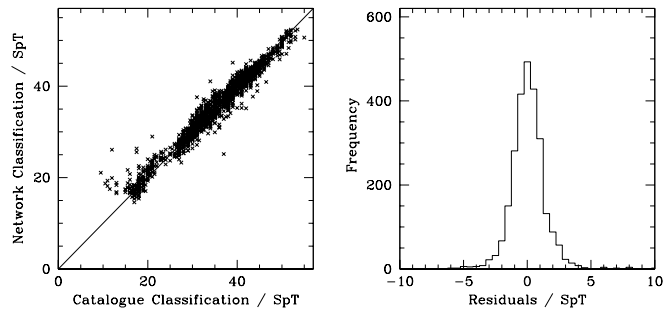,angle=270,width=0.5\textwidth}
}
\caption{Spectral type classification results from a
committee of ten 50:5:1 neural networks
applied to line+continuum spectra.
The left-hand panel is a plot of the committee
classifications against the `true'
classifications listed in the MHD catalogue. The diagonal line is the
locus of points for which the committee classifications equal the catalogue
classifications, and is drawn to guide the eye. Note that even
a perfect classifier would not give results 
exactly on this line on account of noise in the spectra and uncertainty
in the `true' classifications. The right-hand panel is a histogram
of the classification residuals, $\cmean - G^p$, where $\cmean$ is the
committee classification and $G^p$ the
catalogue classification of the $p^{th}$ spectrum. 
The classification error is $\sigma_{\rm{68}} = 1.07$ SpT.}
\label{anal01}
\end{figure}
Figure~\ref{anal01} shows classification results using a committee of
ten networks applied to the 50-component line+continuum PCA spectra.
Each network has a 50:5:1 architecture, indicating 50 nodes in the
input layer, 5 nodes in the hidden layer and a single output (numbers
exclude bias nodes).  The average classification error is
$\sigma_{\rm{68}} = 1.07$ SpT, which has an associated uncertainty of
$\varepsilon = 0.02$ SpT, i.e.\ $\sigma_{\rm{68}} = 1.07 \pm 0.02$
SpT. $\sigma_{\rm{rms}} = 1.41$ SpT, indicating that the tails of the
distribution of the residuals are `heavier' than we would get if the
distribution were Gaussian.

\begin{figure}
\centerline{
\psfig{figure=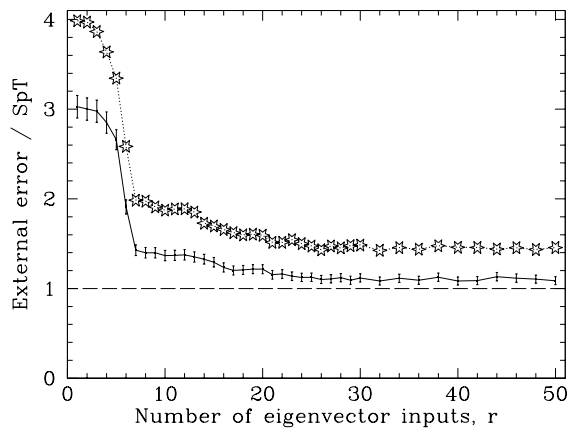,angle=270,width=0.45\textwidth}
}
\caption{Variation of network classification (external) error, as a function
of number of PCA inputs to the neural network. The solid line is
\sig68\ and the dotted line is \sigrms. The number of
inputs, $r$, is the number of principal components used to represent
each spectrum. The error bars on the lower curve are $3 \times \varepsilon$
errors (see equation~\ref{err2}). 
Statistically speaking, there is a significant drop in 
the classification error from $r=1$ to $r\approx25$, followed
by a barely significant decrease in error to $r=50$.
}
\label{err02}
\end{figure}

The optimality criterion for reconstructing spectra
(section~\ref{PCA_comp}) specified that 25 principal components would
give an optimal reconstrutcion.  However, this figure was achieved
without any reference to how we would subsequently use the spectra.  A
direct evaluation of the number of principal components required is
shown in Figure~\ref{err02}, which summarizes the performance of
networks with an $r$:5:1 architecture for a range of $r$.  The
behaviour is well anti-correlated with the behaviour of the
reconstruction quality, $R$, in Figure~\ref{reconerr_whole}.  This is
what we would expect and demonstrates that the network is making best
use of the information given to it.

\begin{figure}
\centerline{
\psfig{figure=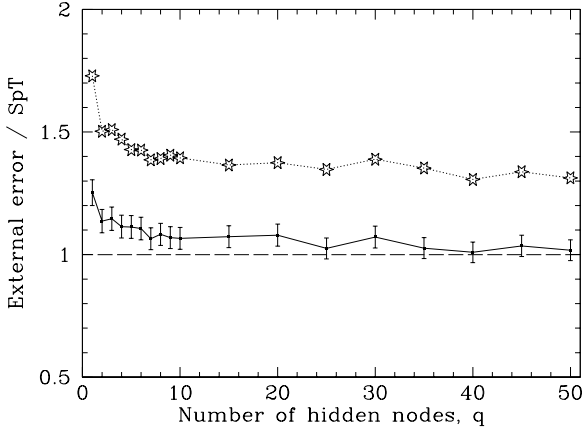,angle=270,width=0.45\textwidth}
}
\caption{Variation of the network classification (external) 
error as a function of
the number of hidden nodes, $q$, in a 50:$q$:1 neural network.
For each value of $q$ a single neural network was trained using
line+continuum spectra and its performance assessed by measuring
\sig68\ (solid line) and \sigrms\ (dashed line).  
The error bars are $3 \times \varepsilon$, where $\varepsilon$ is the
statistical uncertainty in \sig68.
}
\label{err03}
\end{figure}

As the number of hidden nodes in the network increases, so does its
ability to accurately model increasingly complex input--output
functions.  Figure~\ref{err03} shows how the classification errors
vary with the number of hidden nodes
These results show that as $q$ is increased to about seven, there is
an improvement in classification performance, but beyond this there is
no statistically significant improvement. (Note that the neural
network with only one hidden node can only linearly discriminate
between spectral types, which explains the sharp decrease in
classification error between $q=1$ and $q=2$.)  A network with $q=250$,
gave $\sigma_{\rm{68}} = 1.00 \pm 0.01$ SpT, which is a small increase over
$q \approx 5$.  While it is theoretically true that a sufficiently
large number of hidden nodes will increase performance (Hornick,
Stinchcombe \& White 1990)\nocite{hornick_90a}, the number of hidden
nodes required would be inhibitively large, both on account of
training time and over-fitting the data.  The solution is to use an
additional hidden layer.  This can provide increased complexity with a
smaller total number of weights.  A committee of ten 50:5:5:1 neural
networks gave a classification error of $\sigma_{\rm{68}} = 0.88 \pm
0.01$ SpT, which is significantly better than the results with only
one hidden layer.

\begin{table*}
\begin{minipage}{12cm}
\begin{center}
\caption{Summary of the spectral type classification results 
using committees of ten networks for a variety of architectures.
The error measures in columns 3--6 are defined in section~\ref{err_meas}.
}
\begin{tabular}{ccccccc}\hline
Network       & Spectral        & \sig68  & $\varepsilon$  & \sigrms     & $\sigma_{\rm{int}}$ & No. of  \\
Architecture  &  Format         &  (SpT)  &    (SpT)       &  (SpT)      & (SpT)  &  Weights \\ \hline
25:3:1        & line+continuum      &  1.11   &  $\pm$0.015    &  1.43       &  0.19  &   82     \\
25:5:1        & line+continuum      &  1.11   &  $\pm$0.015    &  1.46       &  0.18  &  136      \\
50:5:1        & line+continuum      &  1.07   &  $\pm$0.015    &  1.41       &  0.24  &  261     \\
25:5:5:1      & line+continuum      &  0.86   &  $\pm$0.012    &  1.16       &  0.33  &  166     \\
50:5:5:1      & line+continuum      &  0.88   &  $\pm$0.012    &  1.16       &  0.47  &  291     \\
820:5:5:1     & line+continuum      &  0.82   &  $\pm$0.011    &  1.18       &  0.15  & 4141     \\
25:3:1        & line-only       &  1.09   &  $\pm$0.015    &  1.43       &  0.28  &   82     \\
25:5:1        & line-only       &  1.04   &  $\pm$0.015    &  1.37       &  0.34  &  136     \\
50:5:1        & line-only       &  1.03   &  $\pm$0.014    &  1.35       &  0.41  &  261     \\
25:5:5:1      & line-only       &  0.82   &  $\pm$0.011    &  1.09       &  0.36  &  166     \\ 
50:5:5:1      & line-only       &  0.86   &  $\pm$0.012    &  1.15       &  0.56  &  291     \\ 
25:25:25:1    & line-only       &  0.86   &  $\pm$0.012    &  1.15       &  0.53  &  1326    \\
MHD class.    & photographic    &         &                &             &        &          \\
(Houk)        & plates          &  0.63   &  --            &  --         &  0.44  &  ?       \\ \hline   
\end{tabular}
\label{tempsum}
\end{center}
\end{minipage}
\end{table*}
Table~\ref{tempsum} is a summary of the spectral type classification
results obtained with a range of network architectures. The most
significant result is that smaller external errors (\sig68\ and
\sigrms) are obtained with two hidden layers compared with one hidden
layer. Interestingly, the internal error is smaller when the total
number of weights is smaller, presumably because the minimum of the
error function is easier to locate consistently when the
dimensionality of the space is lower. This is one reason for using PCA
to compress the spectra, as it results in a network with fewer
weights.

Our results also show a very small improvememnt in classification
performance when using line-only spectra. This is perhaps to be
expected as the continuum is more contaminated by effects such as
interstellar reddening and non-linear photographic response.  It is
interesting that von Hippel et~al.\ (1994) (Paper
I)\nocite{vonhippel_94a} obtained worse results with line-only
spectra.  This is probably because they used spectra at half the
resolution, resulting in the loss of some line-information (whereas
the continuum would hardly be affected).

\begin{figure}
\centerline{
\psfig{figure=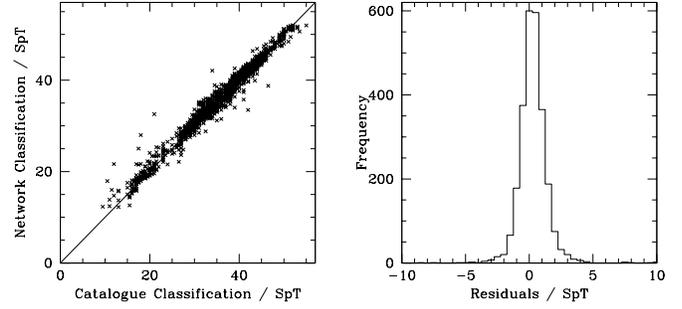,angle=270,width=0.5\textwidth}
}
\caption{Best spectral type results ($\sigma_{\rm{68}} = 0.82$ SpT),
obtained with a committee of ten 25:5:5:1 networks with line-only
spectra.  }
\label{anal06}
\end{figure}

The best results are plotted in Figure~\ref{anal06}.
Table~\ref{anal06_tab} lists the \sig68\ error achieved by each neural
network in this committee and confirms empirically that the error
obtained by the ten networks acting as a committee is lower than any
one of the networks acting alone. Figure~\ref{multi06a} shows
graphically the degree of reproducibility of results.  The committee
result of $\sigma_{\rm{68}} = 0.82$ compares quite favourably with the
error in the catalogue classifications themselves, estimated to be
$\sigma_{\rm{68}} = 0.63$ SpT (N.\ Houk, private communication, 1995),
which represents a lower limit on the precision we can achieve.
Larger networks did not improve performance further.

\begin{table}
\begin{center}
\caption{Spectral type classification errors for each member a
committee (25:5:5:1 networks on line-only spectra).  The committee
error is lower than the error achieved by any one of its members
acting alone. The dispersion of results for individual network is
typical of the other committees used in the spectral type problem.}
\begin{tabular}{|c|c|c|}\hline
Network  &  \sig68  &  \sigrms \\ \hline
1   & \hspace{2ex}0.90\hspace{2ex}  & \hspace{2ex}1.21\hspace{2ex}  \\
2   & \hspace{2ex}0.86\hspace{2ex}  & \hspace{2ex}1.23\hspace{2ex}  \\
3   & \hspace{2ex}0.87\hspace{2ex}  & \hspace{2ex}1.09\hspace{2ex}  \\
4   & \hspace{2ex}0.89\hspace{2ex}  & \hspace{2ex}1.27\hspace{2ex}  \\
5   & \hspace{2ex}0.93\hspace{2ex}  & \hspace{2ex}1.24\hspace{2ex}  \\
6   & \hspace{2ex}0.91\hspace{2ex}  & \hspace{2ex}1.21\hspace{2ex}  \\
7   & \hspace{2ex}0.87\hspace{2ex}  & \hspace{2ex}1.17\hspace{2ex}  \\
8   & \hspace{2ex}0.95\hspace{2ex}  & \hspace{2ex}1.31\hspace{2ex}  \\
9   & \hspace{2ex}0.89\hspace{2ex}  & \hspace{2ex}1.13\hspace{2ex}  \\
10  & \hspace{2ex}0.87\hspace{2ex}  & \hspace{2ex}1.17\hspace{2ex}  \\ \hline
{\bf Committee}  &  \hspace{2ex}{\bf 0.82}\hspace{2ex}  
          &  \hspace{2ex}{\bf 1.09}\hspace{2ex} \\ \hline
${\mbox{\boldmath $\sigma_{\rm{int}}$}}$  & \multicolumn{2}{c|}{\bf 0.36} \\ \hline
\end{tabular}
\label{anal06_tab}
\end{center}
\end{table}

\begin{figure*}
\begin{minipage}{14cm}
\centerline{
}
\caption{(This figure is provided as a separate GIF file.)
Performance of the first five members of the committee
networks used to produce the results in Figure~\ref{anal06}.  The
left-hand figures show how the neural network error drops with
increasing iteration number.  The dashed line shows the error on the
training set and the solid line the error on the test set.  Note that
the network error, as measured on the test set during training,
briefly {\it increases} early on in the training for the first and
third networks. This demonstrates that we should not stop training the
instant that the error on the test set rises. It is also interesting
that most of the training takes place in the first few iterations.}
\label{multi06a}
\end{minipage}
\end{figure*}

To prove that the PCA was not limiting the performance of the neural
network classifiers, we trained a committee of neural networks on the
original (non-PCA) 820-bin spectra. The mean classification error of
$\sigma_{\rm{68}} = 0.82$ SpT shown in Table~\ref{tempsum} is no
better than the PCA-input results for comparable numbers of hidden
nodes, confirming that the PCA compression has not resulted in the
loss of any classification-significant information.  Strictly
speaking, this 820:5:5:1 network has too many weights to be
well-determined by the data (4141 weights (unknowns) vs.\ 2500 spectra
(equations)), so it may be surprising that such a network can
generalize.  However, due to correlations between the spectral
features, some input weights will be correlated, effectively reducing
the number of parameters which must be determined by the data.
Indeed, the Principal Components Analysis showed that the effective
dimensionality of the spectra is only about 25.

\begin{figure}
\centerline{
\psfig{figure=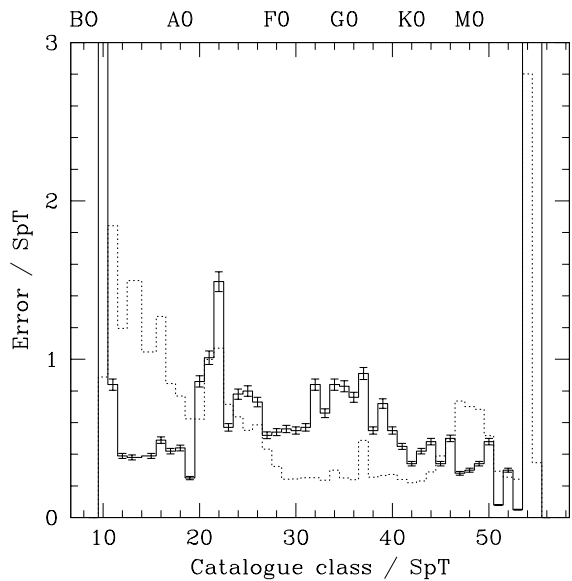,angle=0,width=0.45\textwidth}
}
\caption{Neural network classification errors as a function of
spectral type. The solid line is the external error, \sig68, and
the dotted line the internal error, \sigint.  The error bars on the
external error histogram are $3 \times \varepsilon$ errors for each
spectral type bin, showing that the differences are significant. These
results are from the committee of ten 25:5:5:1 neural networks applied
to line+continuum spectra shown in Table~\ref{tempsum}.}
\label{serr05_p1}
\end{figure}
The internal and external error measures we have been using are
averages over all spectral types.  Figure~\ref{serr05_p1} shows that
\sig68\ and \sigint\ vary considerably as a function of spectral type.
This is influenced by the frequency distribution of spectral types: As
we can see from comparison with Figure~\ref{dist_B}, where there are
relatively few spectra in the training set the classification errors
are correspondingly higher.  This is because the neural network has
been presented with relatively little information about these regions,
and the few spectra are unlikely to give adequate information on the
intra-class variability. Indeed, if we remove the few spectra at the
earliest and latest spectral types, our overall error drops towards
the limit imposed by the training data.

We have also experimented using neural networks in probabilistic mode for
spectral type classification.  A committee of ten such 50:5:5:57
networks applied to the line+continuum spectra gave $\sigma_{\rm{rms}}
= 2.09$ SpT, which is somewhat inferior to continuous output results.
However, the probabilistic approach does offer some advantages,
such as the ability to recognise composite spectra (Weaver
1994)\nocite{weaver_94a}.

\section{Luminosity Class Results}\label{lum_class}

Neural networks were applied to the luminosity class problem in
probabilistic mode (see section~\ref{ann}) using data set B.  The
spectrum is classified as that class for which the output is highest.
We only consider two-hidden layer networks.

The measure of network performance when we have a few discrete classes is
by means of the {\it confusion matrix}. This reports the fraction of
spectra which have been correctly and incorrectly classified for each
class.
\begin{table*}
\begin{minipage}{14cm}
\caption{Confusion matrices for four different committees of ten networks. 
Each confusion matrix lists the percentage of
spectra which have been classified correctly and incorrectly. Thus
 from the top-left matrix we see that the committee correctly
classifies 97.7\% of class Vs as class Vs, but incorrectly
classifies 1.5\% of spectra which are class
V in the catalogue as class III. The
rows of each matrix sum to 100\%. }
\centerline{
\psfig{figure=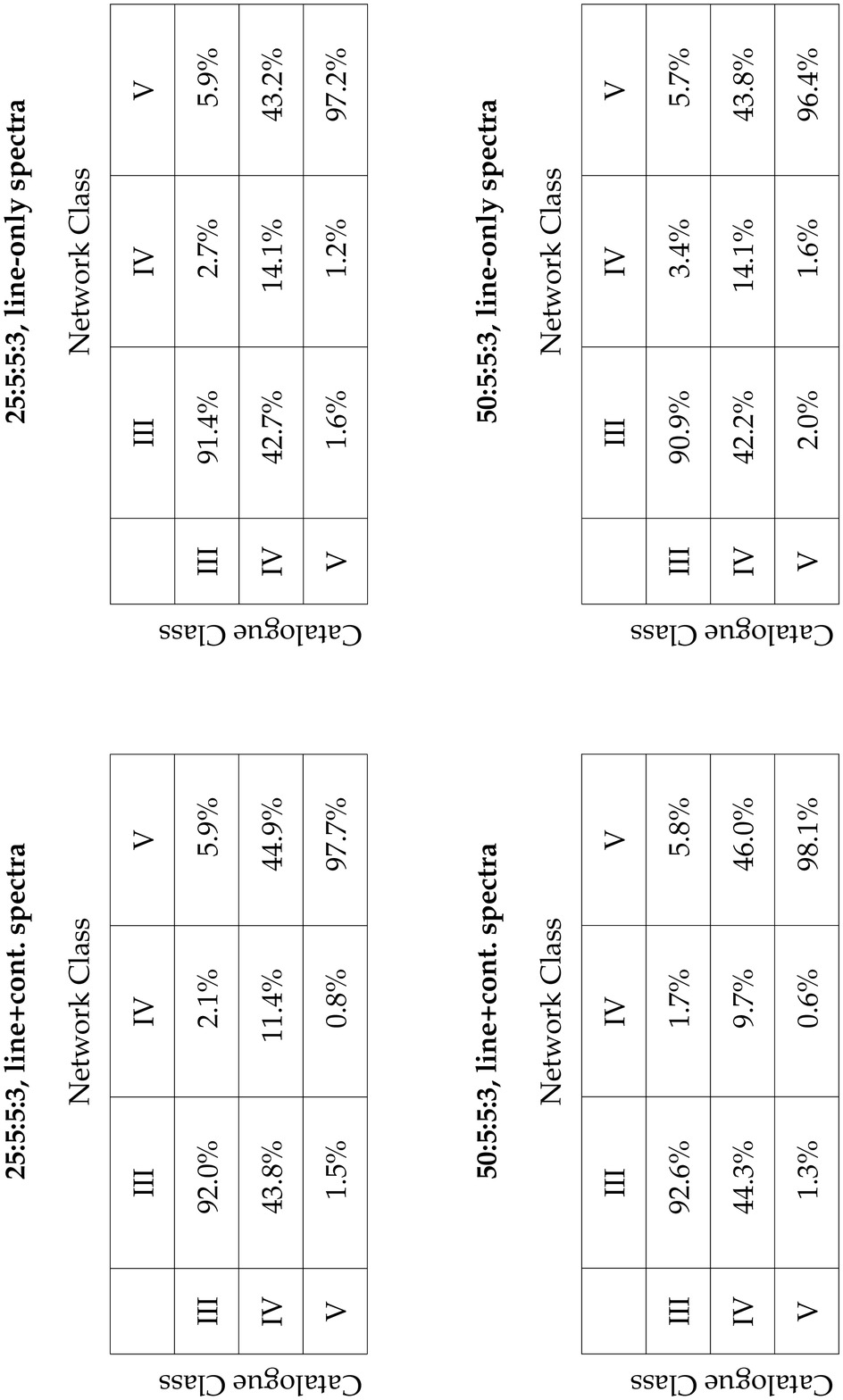,angle=270,width=0.95\textwidth}
}
\label{lumres}
\end{minipage}
\end{table*}
Table~\ref{lumres} compares the results from different committees of
networks, with the four combinations of line-only or line+continuum
spectra represented with 25 or 50 principal components.  We see that
there is little difference in performance between any of these
combinations. This is in agreement with the spectral type
classifications and confirms that most of the luminosity class
information is contained within the first 25 admixture
coefficients. The networks give very good results for classes III and
V (but not class IV), although the better results for class V than
class III may be due to larger fraction of class Vs in the data set
(1.6 times as many).

\begin{figure*}
\begin{minipage}{14cm}
\centerline{
\psfig{figure=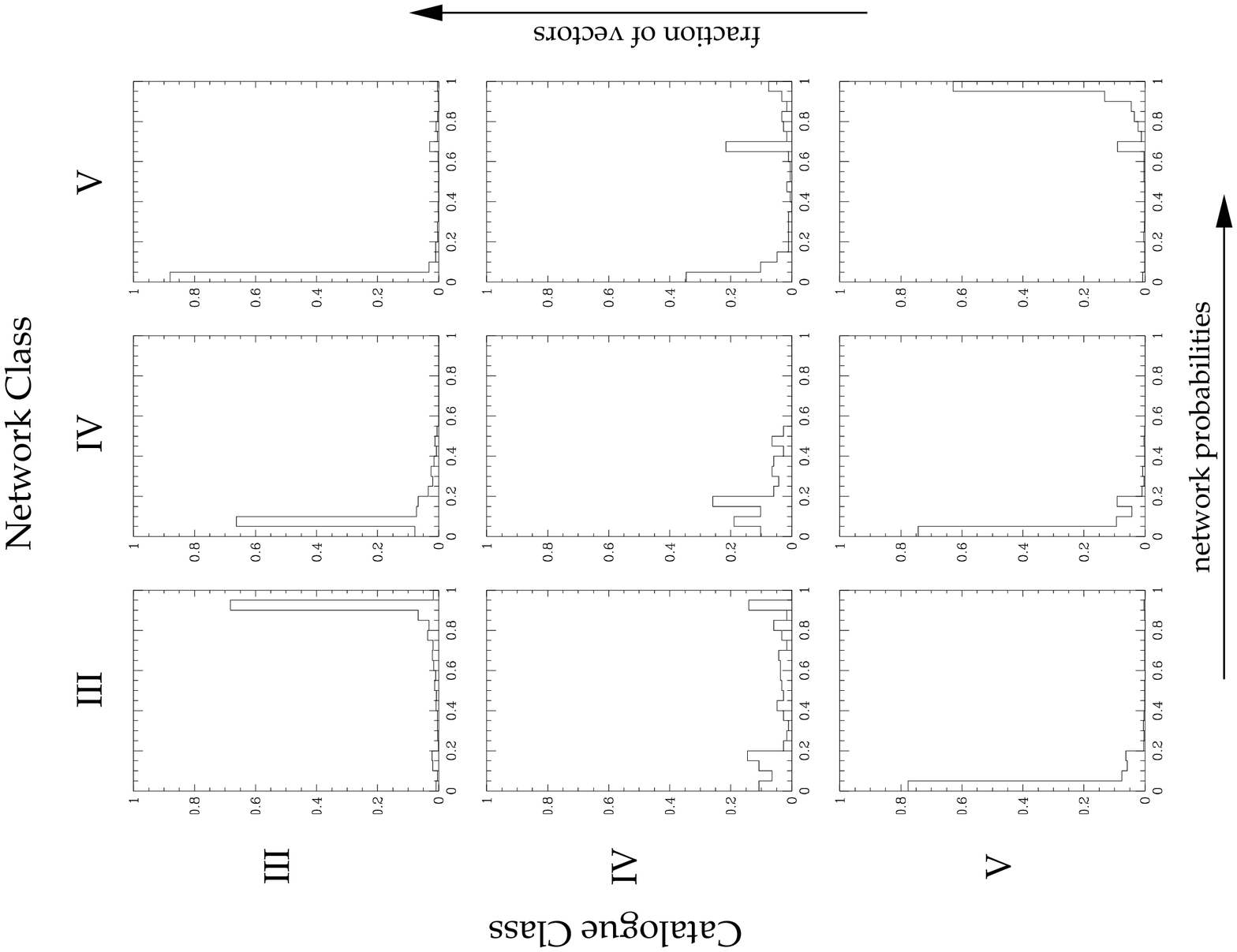,angle=270,width=0.90\textwidth}
}
\caption{Distribution of class probabilities assigned by 
the three outputs of a committee
of ten 50:5:5:3 networks trained on line+continuum spectra.
Each plot is a histogram of the committee probabilities for a certain
network class (column) and a certain catalogue class (row). For
example, the bottom-left hand diagram shows the distribution of the
committee probabilities from the class III output node for spectra
which are catalogue class V, and shows that 
most catalogue class V objects have been assigned a low
probability of being class III. Each plot in a given row
includes the same spectra, and this number has been used to normalize
the fractions for that row.  The total numbers of IIIs, IVs and Vs in
the test data set are 848 IIIs, 185 IVs and 1364
Vs, with almost exactly the same numbers in the training set.
Due to the nature of the sigmoid function
in the neural network we can never achieve 100\% confidence.}
\label{lum08}
\end{minipage}
\end{figure*}
Figure~\ref{lum08} shows the distribution of the probabilities which
the committee assigns for each class.  While most of the class III and
V objects are correctly classified with large confidence, the opposite
is true for class IVs. The nework is not classifying IVs at random
(otherwise we would expect it to classify about 33\% correct).
Rather, the networks have a preference for classifying IVs as either
IIIs or Vs. While the relative paucity of class IVs in the training
set will have some influence, they are not so rare to give such poor
performance. Nor are the IVs lower quality spectra.

Referring back to Figure~\ref{dist_B} we see that there is a fairly
strong correlation between spectral type and luminosity class. Is the
network using spectral type information to produce luminosity
classifications? It would do quite well if it simply classified all
spectra later than about K0 as giants and the rest as dwarfs. (Note
that much of this correlation is real, because the HR diagram is not
uniformly populated.)

\begin{table}
\caption{An `overlap' data set is classified using the committee
of ten 50:5:5:3 networks trained on line+continuum
spectra. This overlap data
set consists of those spectral classes for which the three luminosity
classes are equally represented, which is for spectral types G5/G6,
G6, G6/G8 and G8. This data set consists of
86 IIIs, 83 IVs and 68 Vs. The committee still classifies
IIIs and Vs well and still fails on IVs (cf.\ Table~\ref{lumres}).
}
\centerline{
\psfig{figure=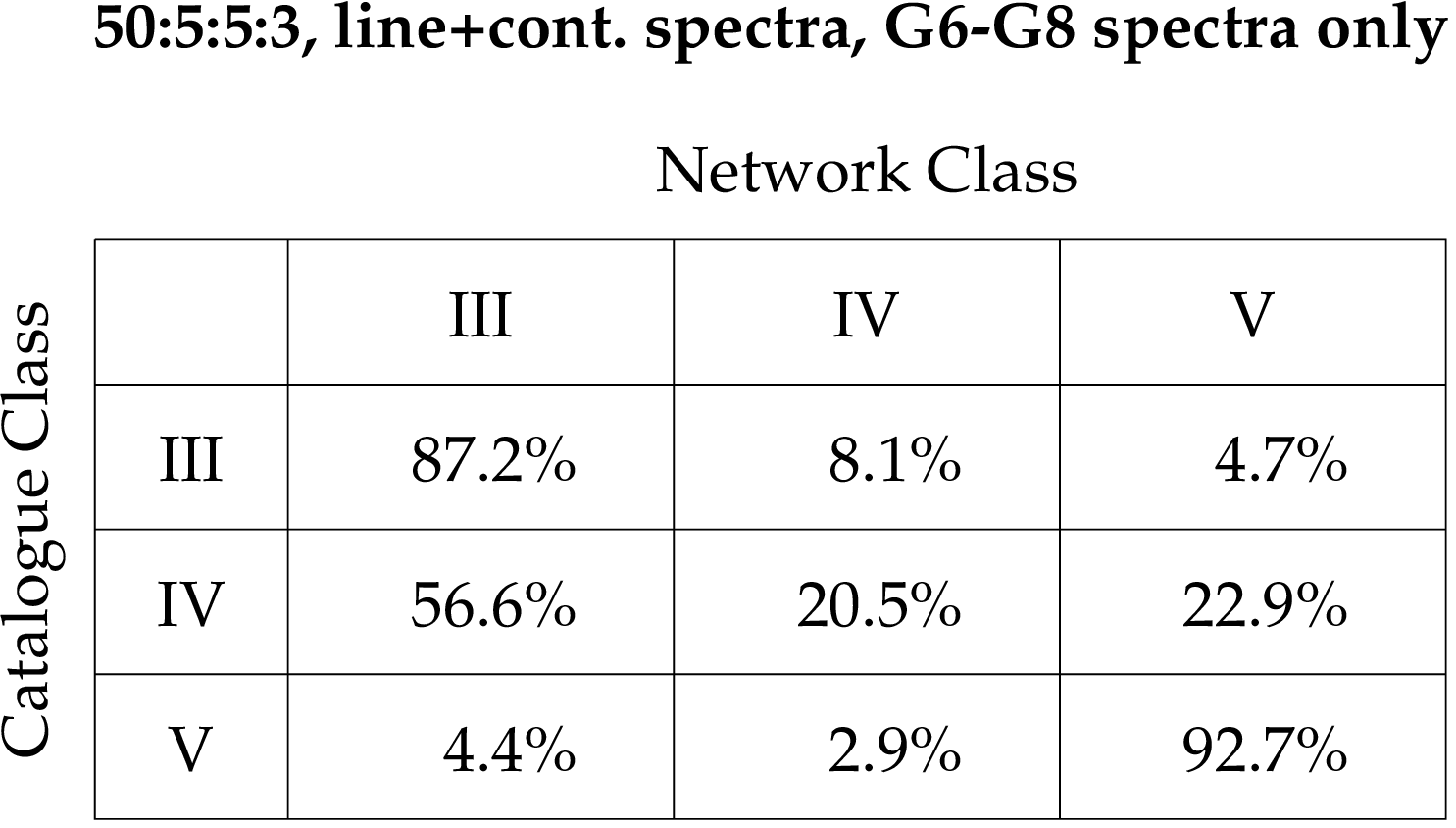,angle=270,width=0.35\textwidth}
}
\label{ovlap08}
\end{table}
To find out what spectral information the networks are using, an
`overlap' data set was created by selecting spectra of classes III, IV
and V in roughly equal numbers for that range of spectral types where
their frequency distributions overlap (around G6).  These spectra were
then classified using a committee previously trained on all the data.
We see from Table~\ref{ovlap08} that the committee still yields good
classifications of classes III and V, for which it cannot
be using spectral type information as there is no correlation between
spectral type and luminosity class over this specifically chosen and
narrow spectral range.  There must, therefore, be independent
luminosity information present in this 50-component reconstruction of
the stellar spectra.  The class IV classifications are still poor.

The failure on class IV spectra implies that, at the resolution of
these spectra, class IV stars are not spectroscopically distinct from
either class III or class V.  Certainly, visual classifiers find it
hard to distinguish class IVs from IIIs and Vs around late G-type
stars (N.~Houk, private communication, 1996). It cannot be due to the
PCA compression as complete spectrum classification gives almost
exactly the same results as shown in Figure~\ref{lumres}.  Problems
with the data reduction, e.g.\ imperfectly registering the spectra in
wavelength, could also contribute.  An alternative explanation is as
follows.  Visual classifiers can focuse on certain lines in a spectrum
and disregard all others. In principal, neural networks can do this
too by altering their weights. However, when there is noise in the
spectrum some inputs will show random correlations with the target
outputs. Thus the network will make a small level of false inference
about the relevance of certain inputs in determining the outputs. With
class IV discrimination the relevance of the few truly important
features may have been washed out this false `noise association'.  A
solution to this problem is to use prior knowledge of which lines are
relevant and train the network only on those features. Another
approach is {\it automatic relevance determination} (MacKay
1995)\nocite{mackay_95a}, which is a Bayesian technique for assessing
the relevance of the inputs using the evidence in the data.

We attempted to use networks with two continuous outputs to tackle the
spectral type and luminosity class problems simulataneously.  However,
the results were inferior, with the best results being \sig68\ = 1.53
SpT (\sigrms\ = 2.02 SpT) for the spectral type and \sig68\ = 0.15
(\sigrms\ = 0.4) luminosity classes (Bailer-Jones 1996).  Due to the
spectral type--luminosity class correlation in the data set, the
network may be unable to adequately separate out luminosity effects
 from temperature ones. This is not helped by the weakness of the
luminosity distiguishing features in this wavelength region. In order
to tackle both problems simultaneously, we may need a more complex
model, and such complexity may not be available with modest-sized
networks.

\section{Summary}

We have produced a system for the automated two-parameter
classification of stellar spectra over a wide range of spectral types
(B2--M7) based on a large ($> 5000$), homogenous set of spectra. We
have shown that we can achieve classification errors of \sig68\ = 0.82
subtypes (\sigrms\ = 1.09 subtypes) over this complete range of
spectral subtypes. This result compares favourably with the intrinsic
errors of \sig68 = 0.63 subtypes in our training data.  Once a neural
network has been trained, its classification results are completely
reproducible. Moreover, the low values of their internal errors
($<0.4$ spectral subtypes) demonstrate that networks can be re-trained
to give sufficiently consistent classifications.

We have achieved correct luminosity class classification for over 95\%
of dwarfs (class V) and giants (class III).  Results for luminosity
class IV spectra were considerably worse.  It is believed
that the data themselves could be a limiting factor and methods for
improving these results were discussed. Despite the correlation in the
data set between spectral type and luminosity class, it was
demonstrated that the neural networks were using luminosity features
to do dwarf-giant discrimination.

Network with two hidden layers performed considerably better ($\approx
0.2$ subtypes) than ones with only one hidden layer.  The best
classification results were achieved by tackling the spectral type and
luminosity class problems separately, using continuous and
probabilistic networks respectively.

We used Principal Components Analysis to compress the spectra by a
factor of over 30 while retaining 96\% of the variance in the data.
It was shown that this compression predominantly removes noise. In
addition the PCA preprocessing reduces the dimensionality of the data
and can be used to filter out bogus spectral features or identify
unusual spectra. However, PCA has the drawback that very weak or rare
features will not be well-reconstructed.  More complex non-linear
preprocessing schemes could no doubt be devised, but the strength of
PCA is its analytic simplicity and its robustness.

The automated classifiers presented in this paper have been used
to produce classifications for several thousand stars which do not
have classifications listed in the MHD catallogue. These
will be presented in a future paper (Bailer-Jones 1998).

\section*{Acknowledgments}

We would like to thank Nancy Houk for kindly loaning us her plate
material.


\end{document}